\documentclass[12pt]{article}

\usepackage[dvips]{graphicx}

\newcommand{\om}{\omega}
\newcommand{\al}{\alpha}
\newcommand{\ep}{\epsilon}

\newcommand{\La}{\Lambda}

\newcommand{\lb}{\lbrack}
\newcommand{\rb}{\rbrack}

\newcommand{\msc}[1]{\mbox{\scriptsize #1}}
\newcommand{\dsp}{\displaystyle}

\newcommand{\cO}{{\cal O}}
\newcommand{\cN}{{\cal N}}

\newcommand{\cS}{{\cal S}}

\newcommand{\tr}{\mbox{Tr}}

\newcommand{\tf}{\tilde{f}}

\newcommand{\tC}{\tilde{C}}

\newcommand{\tN}{\tilde{N}}

\newcommand{\ty}{\tilde{y}}

\newcommand {\eqn}[1]{(\ref{#1})}

\makeatletter
\@addtoreset{equation}{section}

\makeatother

\begin{document}

\begin{titlepage}
 \
 \renewcommand{\thefootnote}{\fnsymbol{footnote}}
 \font\csc=cmcsc10 scaled\magstep1
 {\baselineskip=14pt
 \rightline{
 \vbox{\hbox{hep-th/0305050}
       \hbox{UT-03-14}
       }}}

 \baselineskip=20pt

 \begin{center}
 \centerline{\LARGE  Branches of $\cN=1$ Vacua and Argyres-Douglas Points } 

 \vskip2cm
\noindent{ \large Tohru Eguchi  and Yuji Sugawara} \\
{\sf eguchi@hep-th.phys.s.u-tokyo.ac.jp~,~
sugawara@hep-th.phys.s.u-tokyo.ac.jp}
\bigskip

 \vskip .6 truecm
 {\baselineskip=15pt
 {\it Department of Physics,  Faculty of Science, \\
  University of Tokyo \\
  Hongo 7-3-1, Bunkyo-ku, Tokyo 113-0033, Japan}
 }
\end{center}

\bigskip 

\bigskip

\begin{abstract}

We study the ${\cal N}=1$ version of Argyres-Douglas (AD) points
by making use of the recent developments in understanding 
the dynamics of the chiral sector of $\cN=1$ gauge theories. 
We shall consider $\cN=1$ $U(N)$ gauge theories with an adjoint matter 
and look for the tree-level superpotential $W(x)$ which reproduces the 
$\cN=2$ AD points via the factorization equation 
relating the $\cN=1$ and $\cN=2$ curves.
We find that the following superpotentials generate the $\cN=2$
AD points:\\
(1) $W'(x)=x^N \pm 2\Lambda^N$, \hskip4mm (2) $W'(x)=x^n, \hskip3mm 
N-1\ge n \ge N/2+1$.\\
\indent In case (1) the physics is essentially the same as the 
$\cN=2$ theory even in the presence of the superpotential. 
There seems to be an underlying structure of $N$-reduced KP hierarchy 
in the system. 

Case (2) occurs at the intersection of a number of $\cN=1$ vacua 
with massless monopoles. This branch of vacua is characterized 
by having $s_+=0$ or $s_-=0$ where $s_{\pm}$ denotes the number 
of double roots in $P_N(x)\pm 2\Lambda^N$. 
It is possible to show that the mass gap in fact vanishes at this AD point.
We conjecture that it represents a new class of $\cN=1$
superconformal field theory.

\end{abstract}

\vfill

\setcounter{footnote}{0}
\renewcommand{\thefootnote}{\arabic{footnote}}
\end{titlepage}
\baselineskip 18pt

\section{Introduction}

\indent

Recently, there has been remarkable progress in understanding the
quantum dynamics of a wide class of $\cN=1$ supersymmetric QCD's, 
which share the same field contents as $\cN=2$ SQCD's.    
A series of papers by Dijkgraaf and Vafa \cite{DV1,DV2,DV3} 
has revealed a beautiful correspondence between the effective 
superpotentials in $\cN=1$ SQCD's and the free energies of certain
matrix models. This correspondence 
has been explained based on the arguments 
of geometric transition and topological string theories 
\cite{Vafa,CIV,CV}, which is the B-model version 
of the duality studied in \cite{GV,OV1,OV2}. 
More recently, the correspondence has been proved by means only of field theory
analysis without the help from string theory \cite{DGLVZ,CDSW}. 
Especially, the technology developed by Cachazo, Douglas, Seiberg and 
Witten (CDSW) \cite{CDSW,Seiberg,CSW,CSW2}, based on the anomalous Ward
identity of a generalized Konishi anomaly \cite{Konishi}, 
provides a powerful machinery for the non-perturbative 
analysis on the F-term dynamics of $\cN=1$ SQCD. 
There are a number of subsequent works studying
models with various gauge groups 
and matter contents. A partial list for them is  
\cite{Feng,AO,Witten,BINOR,BFHN,NSW2,AC,CT,KRS}.

In these analyses a crucial role is played by the so-called factorization
equation which relates the $\cN=2$ Seiberg-Witten curve \cite{SW} with the 
reduced $\cN=1$ curve in the presence of massless monopoles.
Let us consider the $\cN=1$ $U(N)$ gauge theory with an adjoint matter field
$\Phi$ with a tree-level superpotential 
\begin{eqnarray}
 \tr\, W(\Phi) \equiv \sum_{m=0}^{n}\, \La^{n-m}\frac{g_m}{m+1}
\tr\, \Phi^{m+1}~,~~
(g_n= 1)~,
\label{tree sp}
\end{eqnarray}
where $\La$ is the dynamical scale parameter.
The factorization equation for the Seiberg-Witten (SW) curve is given by
\begin{eqnarray}
&&\Sigma_{N=2}:\hskip3mm 
y_{N=2}^2=P_N(x)^2-4\Lambda^{2N}=H_{N-n}(x)^2\Big(W'(x)^2+f_{n-1}(x)\Big),
\label{factor1}
\end{eqnarray}
and the $\cN=1$ curve is defined by \cite{CIV,CV} 
\begin{eqnarray}
&&\Sigma_{N=1}: \hskip3mm y_{N=1}^2=W'(x)^2+f_{n-1}(x).
\label{factor2}\end{eqnarray}
Here $P_N(x)$ denotes the standard
characteristic polynomial 
\begin{equation}
P_N(x)=\mbox{Det}\big(x\cdot I-\Phi(x)\big)=x^N+\sum_{i=0}^{N-1}s_ix^{i}
\end{equation}
and $H_{N-n}(x),\,f_{n-1}(x)$ are polynomials of degree $N-n$ and $n-1$, 
respectively.

In the $\cN=1$ theory the eigenvalues of the adjoint field $\Phi$ are 
distributed at
the zeroes of the superpotential $W'(x)$ with multiplicities $N_i, 
\,(i=1,\dots,n)$
and the $U(N)$ symmetry is classically 
broken to $\prod_{i=1}^n U(N_i)$. $N-n$ double 
zeroes
of (\ref{factor1}) implies the existence of $N-n$ massless monopoles in the 
system and the
condensation of monopoles generates the mass gap and confines the gauge 
theories $U(N_i), (i=1,\dots,n)$. Due to confinement one is left with 
$U(1)^n$ unbroken gauge symmetry.

Given the tree-level superpotential $W'(x)$ the factorization equation 
(\ref{factor1}),(\ref{factor2}) completely determines all the parameters of 
the 
polynomials $P_N(x),H_{N-n}(x)$ and $f_{n-1}(x)$. In fact there are $2N$ 
equations
for $N+(N-n)+n=2N$ unknowns (we are putting the coefficients of the
highest order terms of $P_N,H_{N-n},W'$ to 1) and all the unknown parameters 
are expressed in terms of those of the superpotential $W'$.

Thus we have the structure of a vacuum bundle whose base is given by the 
parameter space of the superpotential and the fiber consists of 
the solutions of the factorization equation \cite{Ferrari,CSW}.
Solutions of the equation are partially classified by the integers $s_+\,,s_-$
which describe how the
double zeroes of ${H_{N-n}}^2$ in (\ref{factor1}) are shared by the factors
$P_N\pm 2\Lambda^N$.
$s_+\,,s_-$ denote the degrees of double roots of the polynomials
$P_N+2\Lambda^N,\,P_N-2\Lambda^N$, respectively \cite{CSW}.   
Obviously $s_+,s_-$ must obey the condition
\begin{equation}
s_+ + s_-=N-n.
\end{equation}

Let us now recall some properties of the Argyres-Douglas points where the 
$\cN=2$ gauge theory exhibits the $\cN=2$ superconformal symmetry 
\cite{AD,APSW,EHIY}. 
It is known that $\cN=2$ AD points fall into the A-D-E classification 
corresponding to the A-D-E degeneration of ALE space in string compactification
over Calabi-Yau manifold which is an ALE space fibered 
over ${\bf CP}^1$ \cite{GVW,ShV,GKP}.
 In the case of $A_{N-1}$ type ($N\ge 3$), AD point is obtained simply by 
adjusting the moduli
parameters of the characteristic polynomial $P_N(x)$ as
\begin{equation}
s_i=0~,~~(i=1,\dots,N-1)~, ~~~ s_0=\pm 2\Lambda^N.
\end{equation}
Thus
\begin{equation}
P_N(x)=x^N\pm 2\Lambda^N~,
\end{equation}
and therefore
\begin{eqnarray}
&&y_{N=2}^2=(P_N(x)+2\Lambda^N)(P_N(x)-2\Lambda^N)=x^N(x^N\pm 4\Lambda^N)
\label{AD}\\
&&\hskip1cm \approx x^N  \hskip3mm \mbox{at} \hskip3mm x\approx 0~.
\nonumber
\end{eqnarray}
This describes a genus $N-1$ Riemann surface with a half of its dual $A,\,B$ 
cycles being degenerate.   

In this paper we look for a superpotential $W(x)$ of $\cN=1$ theory
which reproduces the $\cN=2$
AD points (\ref{AD}) when substituted into the factorization equation 
(\ref{factor1}). We find two classes of superpotentials:
\begin{eqnarray}
&&(1)\hskip3mm  W'(x)=x^N\pm 2\Lambda^N\\
&&(2) \hskip3mm W'(x)=x^n, \hskip3mm N-1\ge n \ge {N\over 2}+1
\end{eqnarray}
These superpotentials realize the $\cN=1$ analogues of Argyres-Douglas 
points. 

In the case (1) $n=N$ and there are no massless monopoles and the system
has no mass gap. 
Physics is thus essentially the same as the 
$\cN=2$ theory even in the presence of the superpotential. 
We find some evidence of an 
underlying structure of the $N$-reduced KP hierarchy.
Since $W(x)=x^{N+1}/(N+1)\pm 2\Lambda^N x$,
we obtain a generalized Kontsevich model \cite{GKM}
(with the source term proportional to
an identity matrix) for the calculation of effective superpotential
of the $\cN=1$ theory.

On the other hand, the case (2) occurs at the intersection of 
${}_{n}C_{N-n}$ of $\cN=1$ vacua with $N-n$ massless monopoles.
This branch of vacua is characterized by having $s_+=0$ or $s_-=0$.
Coalescence of phases of different types of monopoles leads to the existence 
of mutually non-local ones and the mass gap in fact vanishes at the AD 
point. We conjecture that this point represents a novel class of
$\cN=1$ superconformal field theory.

This paper is organized as follows.
In section 2 we treat the case (1) using the CDSW method.
We find some evidence for the $N$-reduced KP hierarchy behind this
system. Especially, it is found that 
chiral operators $\langle \tr\, \Phi^m\rangle$ vanishes at 
$m=N\times \mbox{integer}$ and those with $1-N\le m\le 0$
are identified as the flat coordinates of the system.
In section 3 we focus on monomial 
superpotentials $W'(x)=x^n$ and clarify how they realize the $\cN=1$ AD points.
We also discuss on the case of gauge theories with matter fields $N_f\not =0$.
In section 4 we study the scaling behaviors of their chiral ring operators.
We present some comments in section 5.

~


\section{AD points with $W'(x)=x^N\pm2\Lambda^N$}

\indent

Let us first introduce the machinery of CDSW for the calculation
of chiral operators \cite{CDSW}
\begin{eqnarray}
&&T(x)=\left\langle \tr \, \frac{1}{x-\Phi}\right\rangle~,\\
&&R(x)= -\frac{1}{32\pi^2}\left\langle \tr \, 
\frac{W_{\al}W^{\al}}{x-\Phi}\right\rangle.
\end{eqnarray}
Here $W_{\alpha}$ is the field strength chiral field.
The translation dictionary to the variables of Riemann surfaces is given by
\begin{eqnarray}
&&T(x)={P_N(x)'\over y_{N=2}(x)}={d \over dx}\log\Big(y_{N=2}(x)+P_N(x)\Big),
\label{TRiemann}\\
&&R(x)={1\over 2}\Big(W'(x)-y_{N=1}(x)\Big).\label{RRiemann} 
\end{eqnarray}
$T(x)dx$ and $R(x)dx$ are both interpreted as meromorphic one-forms
and are integrated around suitable cycles on the Riemann surface
$\Sigma_{N=1}$. While $T(x)$ is defined in terms of quantities
only of the SW curve $\Sigma_{N=2}$, 
it is well-defined on $\Sigma_{N=1}$ due to the
factorization equation \eqn{factor1}.

In the semi-classical approximation $T(x)\approx P_N'(x)/P_N(x)$ and thus
its integral around some cycle
\begin{equation}
\oint_{A_i} T(x)dx =N_i
\label{flux}
\end{equation}
counts the number of eigenvalues of $\Phi$ inside the integration contour.
Since each eigenvalue of $\Phi$ corresponds to the location of a fractional
$D3$ brane,
the integral of $T(x)$ measures the amount of the RR-flux passing through 
the integration contour. It is known that the above relation (\ref{flux}) holds
also at the quantum mechanical level.

Expectation values of chiral fields are given by
\begin{equation}
U_r\equiv \left\langle \tr \,\Phi^r\right\rangle=\oint x^r\,T(x)\,dx
\label{Uexp}
\end{equation}
where the integration contour is around $\infty$.
We similarly define
\begin{equation}
S_r\equiv \left\langle \tr\,\Phi^rW_{\alpha}W^{\alpha}
\right\rangle=\oint x^r\,R(x)\,dx
\label{Sexp}
\end{equation}

Now let us consider the case of the superpotential $W'(x)=x^N-2\Lambda^N$
(for definiteness we will fix the sign of the 2nd term).
We find
\begin{equation}
y_{N=1}^2=y_{N=2}^2=x^N(x^N-4\Lambda^N), \hskip3mm H_0(x)=1,\hskip3mm
f_{N-1}(x)=-4\Lambda^N.
\label{AD N=2curve}\end{equation}
Thus the $\cN=1$ curve is the same as the $\cN=2$ and the
physics of the system
is also expected to coincide with that of 
the $\cN=2$ case. In the following we see  evidence that 
$\cN=1$ SUSY in the presence of superpotential $W'(x)=x^N-2\Lambda^N$
is enhanced to $\cN=2$ SUSY in the IR limit $\Lambda\rightarrow \infty$.

For even $N$, the $N$ zeroes of $x^N-4\La^N$ 
may be paired up with cuts, and define
$N/2$ ``A-cycles'' $\tilde{A}_1,\ldots, \tilde{A}_{N/2}$ on $\Sigma_{N=1}$.
We also introduce a small cycle ``$C$'' 
encircling the origin $x=0$. In the case of odd $N$ 
we likewise obtain the $\left\lb N/2\right\rb$ A-cycles
$\tilde{A}_1,\ldots, \tilde{A}_{\lb N/2\rb}$.
The remaining zero creates a cut ending at $x=0$, and we 
denote the corresponding cycle as $\tilde{A}_{\lb N/2 \rb +1}$.
The cycle $C$ can be also defined in a similar manner, 
but it must wind around $x=0$ twice, since it passes through the cut associated
with the cycle $\tilde{A}_{\lb N/2 \rb +1}$. In the following we mostly
concentrate on the $N$=even case for the ease of presentation.

At the AD point $T(x)$ is given by
\begin{eqnarray}
T(x) = \frac{P_N'(x)}{y_{N=2}(x)} = 
\begin{array}{ll}
\dsp  \frac{2k x^{k-1}}{\sqrt{x^{2k}-4\La^{2k}}}& ~~ (N=2k) 
\end{array}
\end{eqnarray}
and we obtain the following ``flux distributions'' 
\begin{eqnarray}
&& \oint_{\tilde{A}_i} T(x)dx =2~,~(i=1,\ldots,N/2)~,~~~
\oint_{C} T(x) dx =0~. 
\label{flux even N} 
\end{eqnarray}

In the following we consider the IR limit $\La\, \rightarrow \, \infty$,
or equivalently, focus on the dynamics at the energy scales much smaller
than $\La$. In this limit the cycles $\tilde{A}_i$ become large and move out to
$\infty$, and do not influence the IR physics.
It is hence natural to expect that the quantities relevant for the IR
physics are captured by the integrals around the small contour $C$  which 
stays near the singular point as is discussed in \cite{AD}.
We consider the expectation values of the chiral fields
\begin{eqnarray}
&& U_r=\oint_{C} x^r \,T(x) \,dx,
\label{U AD} \\
&& S_r= \oint_{C} x^r \,R(x) \,dx
\label{S AD}
\end{eqnarray}
and derive their scaling behaviors.
Strictly at the AD point $W'(x)=x^N-2\Lambda^N$ one can show that 
\begin{eqnarray}
U_r =0~,~~~
S_r =0\hskip5mm \mbox{all } \hskip3mm r,
\end{eqnarray}
which is consistent with the scaling invariance.

 Now we consider a small perturbation away from the AD point
\begin{eqnarray}
W'(x)=P_N(x)=x^N-2\La^N-\sum_{m=0}^{N-1}g_m x^m \La^{N-m}~,
\label{mu deform 2}
\end{eqnarray}
and compute expectation values of various chiral operators.
Scaling dimensions of the coupling constants $\{g_m\}$ and chiral fields
$\{U_r,S_r\}$ may be easily evaluated in the following manner:
One applies a scale transformation
\begin{equation}
x\rightarrow \rho^{\gamma}x, \hskip4mm g_m\rightarrow \rho^{\gamma(N-m)}g_m, 
\hskip3mm (m=0,\dots,N-1)
\end{equation}
which transforms $(x^N-\sum_m g_mx^m\La^{N-m})$ 
into $\rho^{\gamma N}(x^N-\sum_m g_mx^m\La^{N-m}$).
Then by using
\begin{equation}
U_r\approx \oint_C x^r {(x^N-\sum_mg_mx^m\La^{N-m})'\over 
\sqrt{x^N-\sum_mg_mx^m\La^{N-m}}} \, dx
\end{equation}
we find
\begin{equation}
U_r(\{\rho^{\gamma(N-m)}g_m\})\approx \rho^{\gamma(r+N/2)}U_r(\{g_m\})
\end{equation}
Thus the chiral field $U_r$ has the scaling dimension $\Delta=\gamma(r+N/2)$.
The value of $\gamma$ is fixed if we assume that
the field $\tr \, \Phi$
has no anomalous dimension. This is the case of $\cN=2$ SUSY when the field
$\Phi$ belongs to the $\cN=2$ vector multiplet. 
By imposing $\Delta(U_1)=1$ we find
$\gamma=1/(1+N/2)$. Therefore we find
\begin{eqnarray}
\Delta(g_m)=\frac{2(N-m)}{N+2},\hskip3mm
\Delta(U_r)={N+2r\over N+2},
\hskip3mm \Delta(S_r)={N+2(r+1)\over N+2}.\label{Delta N=2}
\end{eqnarray}
Dimensions of the coupling constants $\{g_m\}$  agree with the
known results \cite{EHIY}. Note that the dimension of the gaugino 
condensation $S_0=\left\langle\tr \,W_{\alpha}W^{\alpha}\right\rangle$ 
is equal to 1 and thus
$S_0$ may be regarded as the scalar component of some chiral field.

Now we would like to make the following observation:
it is easy to see 
\begin{eqnarray}
\Delta(g_m)\le 1~\Longleftrightarrow~
\left\{
\begin{array}{ll}
 m\ge \frac{N}{2}-1&~~(\mbox{for even $N$}), \\
 m\ge \left\lb \frac{N}{2}\right\rb &~~(\mbox{for odd  $N$}).
\end{array}
\right. 
\end{eqnarray}
and conversely
\begin{eqnarray}
\Delta(g_m)> 1~\Longleftrightarrow~
\left\{
\begin{array}{ll}
 m< \frac{N}{2}-1&~~(\mbox{for even $N$}), \\
 m< \left\lb \frac{N}{2}\right\rb &~~(\mbox{for odd  $N$}).
\end{array}
\right.
\end{eqnarray}

It is well-known that if $\Delta(g_m)\le 1$, $g_m$ should correspond to
a coupling constant in the $\cN=2$ superconformal field theory
in 4-dimensions ($\cN=2\,SCFT_4$),
while if $\Delta(g_m)>1$, it corresponds to a 
VEV of dynamical field  (or the moduli)
 (see, for example, \cite{APSW,ShV}).
In the context of string theory,
$\Delta(g_m)\le 1$ is also interpreted as the non-normalizable vertex
operators (or the ``Seiberg states'')
in the Liouville theory, while 
$\Delta(g_m)>1$ corresponds to the normalizable vertex operators
(or the ``anti-Seiberg states'') \cite{GKP,Pelc,Seiberg-L}.

General chiral perturbations of an $\cN=2$ $SCFT_4$ action 
should have the form
\begin{eqnarray}
&& \Delta S = \int d^4x d^2\theta^+d^2\theta^-\, \sum_{m\,:\,\Delta(g_m)
\le 1} 
\,g_m \cO_m~,
\label{N=2 perturbation} \\
&& \Delta(g_m)+\Delta(\cO_m) =2~.  \nonumber 
\end{eqnarray}
By comparing with \eqn{Delta N=2}, we may identify 
the $\cN=2$ chiral operator $\cO_m$ as the one whose lowest component 
is given by 
\begin{eqnarray}
\tr\, \Phi^{m+2-N/2}, \hskip3mm m\ge {N\over 2}-1.
\label{O U}
\end{eqnarray}
On the other hand,
$g_m$ with $\Delta(g_m)>1$ is identified with the VEV of the operator
\begin{eqnarray}
g_m \Longleftrightarrow  
 \,\left\langle\tr \,\Phi^{N/2-m}\right\rangle, \hskip3mm m \le {N\over 2}-1.
\end{eqnarray}
We have thus two equivalent ways of parameterizing 
the ``small phase space'' of $\cN=2$ $SCFT_4$:
by  the relevant coupling constants $g_m$ ($0<\Delta(g_m)\le 1$),
or the VEV's of relevant operators $\cO_m$ ($1<\Delta(\cO_m)<2$),
which are related by the Legendre transformation.

We note that the superpotential $W(x)={1\over N+1}x^N-2\Lambda^Nx-
\sum_m{g_m\over m+1}x^{m+1}\Lambda^{N-m}$ itself in not quasi-homogeneous and
does not have a well-defined scaling behavior. In the IR limit
it will be dominated by the linear term $\Lambda^N x$ and would not
affect the low-energy dynamics in an essential manner. Actually the value 
\begin{equation}
\Delta(g_m)+\Delta(U_{m+1})={3N+2\over N+2}.
\end{equation}
is not far from 3 expected from
the $\cN=1$ superconformal symmetry.

Let us recall the relation (\ref{flux even N}).
RR fluxes are carried by the cuts $\tilde{A_i}$ which are close to
$\infty$ in the IR limit and the small contour $C$ around the origin 
does not carry any flux.
Thus the analysis of the physics near the singularity $x=0$ becomes the same 
as if there 
were no RR-fluxes or superpotential in the system and we expect that
the original $\cN=2$ 
supersymmetry is restored in the IR limit.   

Let us next specialize to the case of perturbation of the
AD point by keeping only $g_0\equiv \mu\not=0$ and setting 
all other couplings $g_m$ to zero. Then the singularity at the origin
is split into $N$-th roots of unity
\begin{equation}
x^N-\mu\Lambda^N=0 \hskip3mm \Longrightarrow \hskip3mm
x_a=e^{2\pi ia/N}\mu^{1/N}\Lambda,
\hskip2mm a=0,\dots,N-1.
\end{equation}
We may introduce contours $C_{a,b}$ encircling around
$x_a$ and $x_b$ for any pair
of $a,b$.

It is then easy to show that
\begin{equation}
\oint x^r \,T(x)\, dx=\oint \,dx {x^r Nx^{N-1}\over\sqrt{-4\Lambda^N
(x^N-\mu\Lambda^N)}}=0, \hskip4mm \mbox{for   } r\equiv 0 \hskip2mm 
\mbox{mod } N.
\label{mod N}\end{equation}
The above formula (\ref{mod N}) holds for basic contours $C_{a,a+1}$ and thus
for any contours obtained by combining them.

Similarly we have
\begin{equation}
\oint x^{r-1} \,R(x)\, dx=0, \hskip4mm \mbox{for   } r\equiv 0 \hskip2mm 
\mbox{mod } N.
\label{mod Nb}\end{equation}
These are the first instances where we observe the structure of mod $N$ 
reduction in our system.

If we compute the effective superpotential for our theory 
following 
Dijkgraaf and Vafa \cite{DV1,DV2,DV3}, we should 
introduce a matrix action $S_M$
\begin{equation}
S_{M}={1\over N+1}\tr\, M^{N+1}-2\Lambda^N \tr\, M,\hskip4mm N\ge 3
\end{equation}
where $M$ is an $\hat{N}\times \hat{N}$ hermitian matrix.
We note that this is exactly the form of a generalized Kontsevich model
with its source matrix $J$ being replaced by an identity matrix \cite{GKM}.
It is well-known that the free-energy of the generalized Kontsevich model 
obeys Virasoro and W constraints and describes the $N$-reduced KP flow.

In our previous work \cite{ES2} we have studied the string compactification on
a singular $CY_3$ manifold defined by $X^N+z_1^2+z_2^2+z_3^2-\mu\Lambda^N=0$
and discussed its relation to the AD points.
We described this system by an $\cN=2$ 
Liouville theory coupled to an $\cN=2$ minimal 
model (at level $N-2$) following \cite{GKP}. 
The parameter $\mu$ corresponds to the coefficient
of the Liouville potential (cosmological constant) term. After eliminating
variables $z_i,i=1,2,3$ the holomorphic three-form is reduced to an one-form
$\Omega=\sqrt{X^N-\mu\Lambda^N}dX$. Thus $\Omega$ has exactly the same form as
$y_{N=1}=\sqrt{x^N-\mu\Lambda^N}$ at the AD point. Note that the
integral of $1/2\cdot y_{N=1}(x)=R(x)$ (modulo total derivative)
gives the gaugino condensation $S_0$ in gauge theory \eqn{S AD}  while
the integral of $\Omega$ gives a central charge 
or mass of some BPS states in string theory.

Let us introduce further deformation parameters of $CY_3$ and consider
\begin{equation} 
\Omega=\sqrt{X^N-\mu\Lambda^N-\sum_{m=1}t_mX^m\Lambda^{n-m}}.
\end{equation}
and set
\begin{equation}
\left. \Omega_m\equiv {\partial \Omega\over \partial t_m}\right|_{t_m=0}.
\end{equation}
We then find the correspondence
\begin{eqnarray}
\oint_{C_{a,b}} \Omega_m &\Longleftrightarrow &\left. 
{\partial\over \partial g_m}\,S_0(C_{a,b})
\right|_{g_m=0}
\\
&=&\mbox{const}\times \,\oint_{C_{a,b}}
{x^m \over \sqrt{x^N-\mu\Lambda^N}}\, dx,\\
&=& \mbox{const}\times\,U_{m+1-N}(C_{a,b})~,
\end{eqnarray}
where we used the notations 
\begin{eqnarray}
&& S_r(C_{a,b}) \equiv \oint_{C_{a,b}} x^r R(x) dx~,~~~
U_r(C_{a,b}) \equiv \oint_{C_{a,b}} x^r T(x) dx~.
\end{eqnarray}

In \cite{ES2} we  have 
interpreted this quantity as the disc amplitude for a boundary
state $|C_{a,b}\rangle$ with an insertion of 
a chiral primary field $\phi_m$ of $\cN=2$
minimal model dressed by a Liouville exponential
\begin{equation}
\int_{C_{a,b}}\Omega_m=\langle 0|\,\phi_m\,e^{p_m\phi}\,|C_{a,b}\rangle
\end{equation}
where $\langle0|$ denotes the Ramond ground state and $\phi$ is the Liouville 
field. Liouville momentum $p_m$ is proportional to $m/N$ due to the charge 
integrality condition. Charge integrality allows more general values of
the momentum $m/N+s, \,s\in {\bf Z}$ which corresponds to the $s$-th
descendant fields.

We notice a mapping between the deformation parameters of $CY_3$
and operators in the chiral ring  
\begin{equation}
\partial /\partial t_m \Longleftrightarrow U_{m+1-N}.
\end{equation}
Thus the primary couplings (flat coordinates) 
map to chiral operators 
with negative
exponents or the $-1$ descendants. This is a characteristic feature of 
integrable systems. (See, for instance, \cite{Losev,EKYY}.)

~

\section{AD Points with $W'(x)=x^n$}

\subsection{Pure Gauge Theory}

\indent

Let us next discuss the AD points realized by the monomial superpotentials
\begin{equation}
W'(x)=x^n,\hskip5mm {N\over 2}+1\le n\le N-1
\label{monomial sp}
\end{equation}
in $U(N)$ gauge theory with an adjoint matter
field. In order to study these theories we 
first go back to the factorization equation (\ref{factor1}).
\begin{equation} 
P_N(x)^2-4\Lambda^{2N}=H_{N-n}(x)^2\Big(W'(x)^2+f_{n-1}(x)\Big),
\label{factorization 2}\end{equation}
Since $P_N(x)+2\La^N$ and $P_N(x)-2\La^N$ cannot share any zeroes, 
we can classify 
the solutions of \eqn{factorization 2} according to how  
 the zeroes of $H_{N-n}(x)$ are distributed into these two factors.
We denote the number of double zeroes in $P_N(x)+2\Lambda^N$,
$P_N(x)-2\Lambda^N$ as $s_+$ and $s_-$, respectively \cite{CSW}. 
We have three different cases
\begin{description}
 \item[(1)] $s_-=0:$\hskip4mm All the zeroes of $H_{N-n}(x)$ are those 
of $P_N(x)+2\La^N$.
 \item[(2)] $s_+=0:$ \hskip4mm All the zeroes of $H_{N-n}(x)$ are those 
of $P_N(x)-2\La^N$.
 \item[(3)] $s_+s_-\not=0:$ \hskip4mm The cases other than (i) and (ii).
\end{description}

We will soon find that the cases (1) or (2) are relevant for the 
$\cN=1$ AD points. Let us first focus on the case (1).
Since $P_N(x)+2\Lambda^N$ must be divisible by $H_{N-n}(x)^2$ in this case,
we have the structure
\begin{eqnarray}
P_N(x)+2\La^N &=& H(x)_{N-n}^2 F_{2n-N}(x) ~, 
\label{factorization 3a} \\
P_N(x)-2\La^N &=& H(x)_{N-n}^2 F_{2n-N}(x) -4\La^N~,
\label{factorization 3b}
\end{eqnarray}
where $F_{2n-N}(x)$ is some polynomial with degree $2n-N (>0)$.
By comparing \eqn{factorization 3a} and \eqn{factorization 3b}
with \eqn{factorization 2},
we immediately find \cite{CSW}
\begin{eqnarray}
W'(x)=H(x)_{N-n}F_{2n-N}(x)~,~~~ f(x)=-4\La^NF_{2n-N}(x).
\label{sol W f 1}
\end{eqnarray}
Therefore all the zeroes of the superpotential $W'(x)$ are shared by the
polynomials $H_{N-n}(x)$ and $F_{2n-N}(x)$.
In the generic case we have a superpotential with zeroes 
$\{a_i; \,i=1,\dots,n\}$ which are all distinct
\begin{eqnarray}
W'(x)=x^n+\sum_{m=0}^{n-1}g_mx^m\La^{n-m}= \prod_{i=1}^{n}(x-a_i)~, ~~~ a_i
\neq a_j~~(i\neq j)~
\label{W deform}
\end{eqnarray} 
$H_{N-n}(x)$ and $F_{2n-N}(x)$ split these zeroes into two groups.
Let us use the notation $\{p_i;\,i=1,\dots,N-n\}$ for
the zeroes of $H_{N-n}(x)$ and $\{q_i;\,i=1,\dots,2n-N\}$
for those of $F_{2n-N}(x)$ 
\begin{equation}
H_{N-n}(x)=\prod_{i=1}^{N-n}(x-p_i),\hskip4mm F_{2n-N}(x)=\prod_{j=1}^{2n-N}
(x-q_j).
\end{equation}
As a set
\begin{equation}
\{p_1,p_2,\dots,p_{N-n}\}\cup\{q_1,q_2,\dots,q_{2n-N}\}
=\{a_1,a_2,\dots,a_n\}.
\end{equation}
Therefore in general there are 
\begin{equation}
L\equiv{}_nC_{N-n}
\end{equation}
different branches of 
solutions to the factorization equation.
Thus the vacuum bundle of the case (1) has the structure of a 
$L$-fold branched cover over the coupling constant space ${\cal S}$.
Each sheet has a different semi-classical end in the 
limit $\Lambda\rightarrow 0$. As we see by setting $\Lambda=0$ in 
\eqn{factorization 3a},\eqn{factorization 3b},
the gauge symmetry is given by $U(2)^{N-n}\times U(1)^{2n-N}$
at generic points in the semi-classical region. 

By extremizing the effective superpotential of the system it is possible 
to show that the magnitude of the monopole condensation at 
a double point $x=p_i$
is given by \cite{deBO}\footnote{We thank Y. Tachikawa for drawing our 
attention to this formula.}
\begin{equation}
\left\langle \tilde{M}_iM_i\right\rangle=\mbox{const}\times y_{N=1}
(x=p_i)~, \hskip4mm i=1,\dots,N-n~,
\label{dBO}
\end{equation}
where $M_i,\tilde{M}_i$ are the scalar components of the $i$-th
monopole hypermultiplet. If we use the relation (\ref{factor2})
and (\ref{sol W f 1}), we find
\begin{equation}
\left\langle \tilde{M}_iM_i\right\rangle=\mbox{const}'\times 
\prod_{j=1}^{2n-N}(p_i-q_j)^{1/2}~, \hskip3mm i=1,\dots,N-n~.
\label{mass gap}
\end{equation}
At generic points in the parameter space ${\cal S}$, condensates 
$\left\langle \tilde{M}_iM_i\right\rangle,\,i=1,\dots,N-n$ are all non-zero
and generate the mass gap  and confinement in the system.

Let us now consider the special case of monomial superpotentials
$W'(x)=x^n$ ($N/2+1\le n \le N-1$). In this case all the zeroes 
$\{a_i,p_{j},q_{\ell}\}$
are located at the origin and hence
\begin{eqnarray}
&& P_N(x)^2-4\La^{2N}= x^N(x^N-4\La^N)~, \nonumber \\
&& H_{N-n}(x)=x^{N-n}~,~~~f_{n-1}(x)=-4\La^N x^{2n-N}~,
\label{AD 1}
\end{eqnarray}
which is exactly the AD point of $U(N)$ theory (with the highest criticality).
As we approach the AD point at the 
origin $O\equiv \{g_m=0,\,({}^{\forall} m=0, \ldots, n-1)\}$ of the base space 
$\cal S$, the monopole condensations \eqn{mass gap} all vanish and we expect 
to obtain a scale invariant theory (note that we have $n\ge N/2+1$).
As we see in the next section, physical observables exhibit scaling 
behaviors around this point: thus we conjecture that it represents a new class
of $\cN=1$ superconformal field theory.
The $\cN=1$ AD singularity in this case 
is located at the point where $L$ different 
branches of $s_-=0$ vacua collide in the vacuum bundle.
Let us call this AD type singularity as the $AD_+$ point. 
See Fig.1.

The analysis of the case (2) is completely parallel. 
The monomial superpotential \eqn{monomial sp} generates 
a solution  
\begin{eqnarray}
&& P_N(x)^2-4\La^{2N}= x^N(x^N+4\La^N)~, \nonumber \\
&& H(x)_{N-n}=x^{N-n}~,~~~f_{n-1}(x)=4\La^N x^{2n-N}~,
\label{AD 2}
\end{eqnarray}
which we call as the $AD_-$ point.
The vacuum bundle near the $AD_-$ point again becomes an $L$-fold 
branched cover, and it defines a different branch
from that of $AD_+$.

The case (3), on the other hand, is rather difficult to analyze.
Even with a monomial superpotential 
we may obtain solutions of factorization equation  
which does not develop any singularities at $x=0$.

As a simple example, we have found the following solution 
in the case of $N=5$, $n=3$ with $s_+=s_-=1$
(although it is outside of the range of $n$ in \eqn{monomial sp});
\begin{eqnarray}
&& P_5(x)+2\La^5=\left(x+4^{1/5}\La(1+\om)\right)^2
\left(x^3-4^{1/5}\La x^2
+4^{2/5}\La^2(1+\om)x-4^{3/5}\La^3 \om\right) ~, \nonumber \\
&& P_5(x)-2\La^5=\left(x+4^{1/5}\La\om\right)^2\left(x^3+4^{1/5}\La x^2
-4^{2/5}\La^2\om x-4^{3/5}\La^3(1+\om)\right)  ~, \nonumber \\
&& P_5(x)^2-4\La^{10} =\left(x+4^{1/5}\La(1+\om)\right)^2
\left(x+4^{1/5}\La\om\right)^2
\left\{W'(x)^2 +f(x)\right\} ~, \nonumber \\
&& W'(x)^2 +f(x) =
x^6+2\cdot 4^{4/5}\La^4 x^2 + 4\La^5 (-i\sqrt{3})x 
-4\cdot 4^{1/5}\La^6~, 
\end{eqnarray}
where $\om$ is the 3rd root of unity $\om \equiv e^{2\pi i/3}$.
Solutions of this kind do not correspond to 
any IR fixed point. Due to algebraic complexities we have not yet
been able to study the case (3) in detail. 

It seems, however, 
what happens is the following: by adjusting parameters it is possible 
to generate higher order zeros of the function, for instance, 
$P_N(x)+2\Lambda^N$ at some point $x=p_i$. At this point, however, 
$P_N(x)-2\Lambda^N$ can not
vanish and thus the system is effectively reduced to the case of $s_-=0$
with the rank $N$ replaced by some smaller value.
Thus the $\cN=1$ AD points
which appear on branches with $s_+s_-\not=0$ should be of the same type
as those of $AD_{\pm}$ with lower rank gauge group.

It is worthwhile to note that when $n=N-1$, the case (3) does not occur
and hence all the $\cN=1$ vacua of the $U(N)$ gauge theory with an arbitrary
degree-$N$ superpotential $W(x)$ are 
smoothly connected to the $AD_{\pm}$ points.


\subsection{Non-zero Flavor}

\indent

We next consider the system with non-zero flavors $N_f\neq 0$ (i.e. with $N_f$ 
chiral fields $Q_i,\, \tilde{Q}_i,\,i=1,\dots,N_f$ 
in the fundamental and anti-fundamental representations).
For the sake of simplicity we  consider the even flavor case $N_f=2\ell$,
and discuss the case of degenerate bare masses $\{m_i=m,\,i=1,\dots,N_f\}$ 
in order to study higher singular points.
It is convenient to make a constant shift of $\tr \, \Phi$, so that we can 
set  $m=0$ without loss of generality.
The SW curve is then written as \cite{SW2}
\begin{eqnarray}
y_{N=2}^2=C(x)^2 - 4\La^{2(N-\ell)}x^{2\ell}~,
\label{SW curve flavor 1}
\end{eqnarray}
where $C(x)$ is a degree $N$ polynomial of the form 
$C(x)=P_N(x)+\cO(x^{N_f-N})$ ($C(x)=P_N(x)$, in the range $N_f < N$).

Now the higher singularities are achieved by the following steps:
\begin{enumerate}
 \item Choose the moduli parameters so that the curve \eqn{SW curve flavor 1}
possesses the maximal number of massless squarks 
(``highest non-baryonic branch'' \cite{APS2}). 
In other words we require the factorization
\begin{eqnarray}
y_{N=2}^2 = x^{2\ell} \left(\tC(x)^2-4\La^{2\tN}\right)~,
\label{SW curve flavor 2}
\end{eqnarray}
where we have introduced $\tN=N-\ell$ and $\tC(x)$ is of degree $\tN$. 
Physically this means  
that the Higgsing by the condensation of massless squarks 
breaks the gauge symmetry down
to $U(\tN)$.
\item We then tune the parameters in $\tC(x)$ so that the 
``reduced curve'' $\ty_{N=2}^2=\tC(x)^2-4\La^{2\tN}$ describing 
the Coulomb branch of $U(\tN)$ theory
possesses  the higher critical points $\ty^2 \approx x^p$,
which is called the ``class 4 model'' in \cite{EHIY}.
We especially focus on the case of highest criticality
$p=\tN$, which is described by the curve
\begin{eqnarray}
&& y_{N=2}^2=x^{\tN+2\ell}\left(x^{\tN} -4\La^{\tN}\right)~.
\label{curve class 4} \\
&& (\mbox{or} ~~  y_{N=2}^2=x^{\tN+2\ell}\left(x^{\tN} +4\La^{\tN}\right) .)
\nonumber
\end{eqnarray}
\end{enumerate}
It is obvious that we can work out the similar analysis 
as in the previous subsection by replacing  $N$ with $\tN$.
After turning on the monomial superpotential 
in the range of $n$  
\begin{eqnarray}
\frac{\tN}{2} +1 \leq n \leq \tN-1~,
\label{range k 2}
\end{eqnarray}
we obtain the $\cN=1$ AD points which reproduce the SW curve
\eqn{curve class 4}. The vacuum bundle around this point
is again an $L$-fold branched covering space with 
$\dsp L={}_n C_{\tilde{N}-n}$.
We can readily confirm that the squark condensation vanishes at this point 
\begin{eqnarray}
 \left\langle \tilde{Q}_{\tf}Q^{f}\right\rangle=\mbox{const}\times
y_{N=1}(x=-m=0) =0~, ~~~ ({}^{\forall} f, \tf)~,
\end{eqnarray} 
using the formulas presented in \cite{Seiberg,CSW2}.

~


\section{Scaling Behaviors Around the $\cN=1$ AD Points}

\indent

For the sake of simplicity we first consider the case 
of pure Yang-Mills theory.
We introduce the monomial superpotential $W'(x)=x^n$ \
and focus on the $AD_+$ point \eqn{AD 1}. 
The pair of relevant curves is given by 
\begin{eqnarray}
&& \Sigma_{N=2}:~ y_{N=2}^2=x^N(x^N-4\La^N)~, \nonumber \\
&& \Sigma_{N=1}:~ y_{N=1}^2 = x^{2n-N}(x^N-4\La^N)~.
\end{eqnarray}

Our task is to evaluate the scaling behavior
of the observables under the small perturbation
\begin{eqnarray}
W'(x) = x^n + \sum_{m=0}^{n-1}g_m \La^{n-m}x^m~,~~~ |g_m| \ll 1~.
\label{small perturbation}
\end{eqnarray}
We can again apply the same scaling analysis given in
section 2 and find
\begin{eqnarray}
&& U_r(\{\rho^{\Delta(g_m)}g_m\})\approx \rho^{\gamma(N+2r)}U_r(\{g_m\})~, 
\label{scaling U} \\
&& S_r(\{\rho^{\Delta(g_m)}g_m\})\approx 
\rho^{\gamma\left\{2(n+r+1)-N\right\}}S_r(\{g_m\})~, 
\label{scaling S} 
\end{eqnarray} 
where we set the scaling dimensions of the coupling constants as
\begin{eqnarray}
\Delta(g_m) = 2\gamma (n-m)~, ~~~ (i=0,\ldots, m-1)~.
\label{Delta g}
\end{eqnarray}
$\gamma$ is a positive real constant to be determined 
below.

Scaling analysis fixes the ratios of the scaling dimensions
of coupling constants and physical observables, but cannot determine 
the overall normalization constant $\gamma$. To fix it, we 
demand that the effective superpotential 
$\dsp W_{\msc{eff}} \equiv 
\langle \tr\, W(\Phi) \rangle$ should be a marginal operator 
at the AD point\footnote
  {We keep  the ``effective scale parameter'' 
$\La_{\msc{AD}}$ finite, which is defined by 
$\La^{3-(n+1)}\left\langle \tr\, W(\Phi)\right\rangle = \mbox{const.} \times
\La_{\msc{AD}}^3$ in the IR limit $\La\,\rightarrow\,\infty$.
Namely, we rescale as $\La\,\rightarrow\,\La/\ep$,
$g_r\, \rightarrow\, \ep^{\Delta(g_r)}g_r$, where $\Delta(g_r)$
is defined by \eqn{Delta g 2} in the limit $\ep\,\rightarrow\,0$. 
Explicitly, $\La_{\msc{AD}}$ is written as 
$\La_{\msc{AD}} = \mbox{const.}\times \La \mu^{1/\Delta(\mu)}$,  
($\mu \equiv g_0$).
}.
This requirement amounts to $\Delta(U_{n+1})=3$,
which implies $\gamma =3/(N+2n+2)$.
In this way we arrive at the following formulas for scaling dimensions
\begin{eqnarray}
&&\Delta(g_m) = \frac{6(n-m)}{N+2n+2},\hskip3mm 
\Delta(U_r)=\frac{3(N+2r)}{N+2n+2}, \hskip3mm \Delta(S_r)=\frac{3\big(2n-N+2(r+1)\big)}{N+2n+2}.
\label{Delta g 2}
\end{eqnarray}
We note that 
\begin{eqnarray}
&& W_{\msc{eff}} = \sum_{m=0}^{n} g_m U_{m+1},  
\hskip4mm \Delta(g_m)+\Delta(U_{m+1})=3, \hskip5mm (g_n=1).
\end{eqnarray}
A general chiral perturbation 
to the $SCFT_4$ describing the $\cN=1$ AD point is then given by
\begin{eqnarray}
\Delta S = \int d^4x d^2\theta\, \sum_m g_m \cO_{m}~, \hskip5mm 
\left\langle\cO_m\right\rangle=U_{m+1}
\label{N=1 perturbation}
\end{eqnarray}

The deformation \eqn{small perturbation} resolves the singularity at $x=0$, 
creating new cuts inside the circle $C$. See  Fig.2.
These cuts come from the zeroes of $F_{2n-N}(x)$ in the decomposition
\eqn{sol W f 1} and these are paired up to form   
$n-N/2$ new ``$A$-cycles'' on $\Sigma_{N=1}$.
We denote them as $A_i$ $(i=1,\ldots, n-N/2)$.
These cycles together with
$\tilde{A}_j$ $(j=1,\ldots, N/2)$ 
compose the totality of $n$ $A$-cycles on $\Sigma_{N=1}$.
The cycles $\tilde{A}_i$ and $A_i$ are separated by the length scale
$\Lambda$ and describe the physics of different energy scales.
Additional order parameters for the IR dynamics may be given 
by the integrals of $T(x)$ and $R(x)$ along $A_i$ 
(and corresponding $B$-cycles)
as well as $C$ already introduced.

We can show  that even with a perturbed 
superpotential \eqn{small perturbation}
\begin{eqnarray}
\oint_{A_i} T(x) dx =0~,~~~({}^{\forall}i=1,\ldots,n-N/2)~,
\label{no flux AD}
\end{eqnarray}
thus 
\begin{equation}
U_0=0.
\end{equation}


It is easy to generalize our analysis to the non-zero flavor case 
\eqn{curve class 4} by replacing the color $N$ with $\tN\equiv N-\ell$.
The non-trivial point is the existence of another chiral operator
$M(x)^{f}_{\tf}=\left\langle\tilde{Q}_{\tf}(x-\Phi)^{-1}Q^f
\right\rangle$. However, in this case we can readily find that
\begin{eqnarray}
\oint_C x^rM(x)^f_{\tf} dx 
= \delta^f_{\tf} \oint_C x^{r-1} \left(R(x)-R(0)\right) dx~,
\end{eqnarray}
based on the formulas given in \cite{Seiberg,CSW2}.
We thus need not introduce new observables describing 
scaling properties.
We also point out the obvious relation\footnote
      {Here $T(x)$ and $\tilde{T}(x)$ are explicitly written as \cite{CSW2} 
    (see also \cite{NSW})
   $$
T(x) = \frac{d}{dx} \log \left(C(x)+y_{N=2}(x)\right)~,~~~
\tilde{T}(x) = \frac{d}{dx} \log \left(\tilde{C}(x)+\ty_{N=2}(x)\right)
\equiv \frac{\tilde{C}'(x)}{\ty_{N=2}(x)}~.
   $$}
\begin{eqnarray}
T(x)=\tilde{T}(x)+\frac{\ell}{x}~,
\end{eqnarray}
where $\tilde{T}(x)$ is  the counterpart of $T(x)$  for 
the pure gauge $U(\tN)$ theory corresponding to the curve 
$\ty_{N=2}^2=\tC(x)^2-4\La^{2\tN}$ defined in the factorization relation 
\eqn{SW curve flavor 2}.
The scaling formulas \eqn{Delta g 2} likewise hold 
with replacing  $N$ by $\tN$.

~

\section{Summary and Comments}

\indent

In this paper we have proposed a new class of IR fixed points
with $\cN=1$ supersymmetry which occur in a theory with a
monomial superpotential $W'(x)=x^n, \,N/2+1\le n \le N-1$.
They are located at the ramification points in the vacuum bundle 
over the coupling constant space and have been obtained 
by solving $s_-=0$ or $s_+=0$ branches of the factorization equation.
Turning on small perturbations,  
we have analyzed  the scaling behaviors of chiral ring operators.
Our results strongly suggest that these $\cN=1$ AD points define 
a new universality classes of $\cN=1$ $SCFT_4$.   

We have also studied the $\cN=2$ AD points of $A_{N-1}$ type
and confirmed the consistency with the approach of superstring theory 
compactified on the singular CY 3-folds and described 
by an $\cN=2$ minimal model coupled to the $\cN=2$ Liouville theory
\cite{GKP,ES2}.
In particular we have found that higher power
chiral operators in the gauge theory side are identified with the 
gravitational descendants in the Liouville theory. 
We have found some evidence of the structure of the underlying $N$-reduced KP 
hierarchy in this system.

~


Finally we would like to present several comments:

\noindent
{\bf 1.}
One might suppose that the tree level superpotential 
$W(x)\approx x^{n+1}$ with higher values of $n$ 
could not affect the physics of IR region, since the naive power counting 
tells us that it is an irrelevant operator in the IR.
One thus might guess that our $\cN=1$ AD points 
belongs to the same universality class as the usual 
$\cN=2$ AD points. However, as emphasized in \cite{KSS}, 
these superpotentials are the {\em dangerously irrelevant\/}
operators in the theory of renormalization group. 
They behave as irrelevant operators around the UV fixed point
(Gaussian fixed point), but do not necessarily around the IR fixed point 
when they could gain large anomalous dimensions due to the strong 
quantum effects. 
They could modify the IR physics and generate a new universality class.
Our scaling analysis actually suggests this is indeed the case:
The $\cN=1$ AD points belong to the different universality classes 
from the $\cN=2$ one (note that the dimension of the operator 
$\tr \, \Phi^{n+1}$ is equal $\frac{N+2(n+1)}{N+2}$ \eqn{Delta N=2} 
at the $\cN=2$ AD point, which is less than 3 
when $\frac{N}{2}+1\leq n \leq N-1$, 
and thus it is a relevant perturbation).

~

\noindent
{\bf 2.}
We have observed that, when parameters are adjusted near the AD points,
fluxes are localized on the large cycles of order $\Lambda$ 
and hence move out to infinity under 
the IR limit $\La \,\rightarrow\, \infty$. 
This fact is essential for the appearance of non-trivial $SCFT_4$.
Moreover, in the $\cN=2$ case $W'(x)=P_N(x)$,
this phenomenon gives the theoretical justification why we can describe 
the AD points based on the superstring compactified on 
the singular $CY_3$ {\em without flux\/}. 

~

\noindent
{\bf 3.}
It may be worthwhile to point out the similarity of our analysis 
to that of the topological Landau-Ginzburg models 
coupled with two-dimensional gravity \cite{Losev,EKYY}. 
The approaches from  the integrable hierarchy \cite{CM,CMMV,IM}
is  presumably useful in order to uncover the precise relation 
among these models.
It is also interesting to look for the string theory description 
for the general $\cN=1$ AD points as in the case of $\cN=2$ points.

~

\noindent
{\bf 4.}
 We have defined the $\cN=1$ AD points by requiring 
that {\em both\/} of the Riemann surfaces
$(\Sigma_{N=2}, \Sigma_{N=1})$ develop higher isolated singularities 
at the same point. 
If the point $x=0$ is a singular point on
$\Sigma_{N=2}$, but smooth on $\Sigma_{N=1}$,
no interesting IR physics will be obtained.
According to the formula \eqn{dBO}, 
monopole condensation persists if $y_{N=1}\not =0$ at 
$x=0$ and we will be led to the
confinement and mass gap. As is well-known that for $\cN=2$ AD points
one needs $y_{N=2}^2\approx x^p,\,\,p\ge 3$ around the singularity $x=0$. 
We have imposed in this paper a condition $y_{N=1}^2\approx x^q,\,\,q\ge 2$.
In the exceptional case $y_{N=1}^2\approx x$ (where $\Sigma_{N=1}$ has 
no singularity), we still have a gap-less theory. However, 
it is easily found that all the VEV's of chiral operators 
vanish even under the perturbed superpotential \eqn{small perturbation}. 

~

\noindent
{\bf Note added :}
After the completion of this work we became aware of the papers
\cite{TY,GVY,AGK}, in which $\cN=1$ extensions of Argyres-Douglas 
superconformal theory have also been discussed. 
We would like to thank S. Terashima,  A. Gorsky and K. Konishi 
for bringing our attention to these references.


~


\section*{Acknowledgements}
\indent

The research of T. E. and Y. S. is partially  supported by 
Japanese Ministry of Education, 
Culture, Sports, Science and Technology.



\begin{figure}
\centering
\begin{minipage}{\linewidth}

\includegraphics[width=\linewidth]{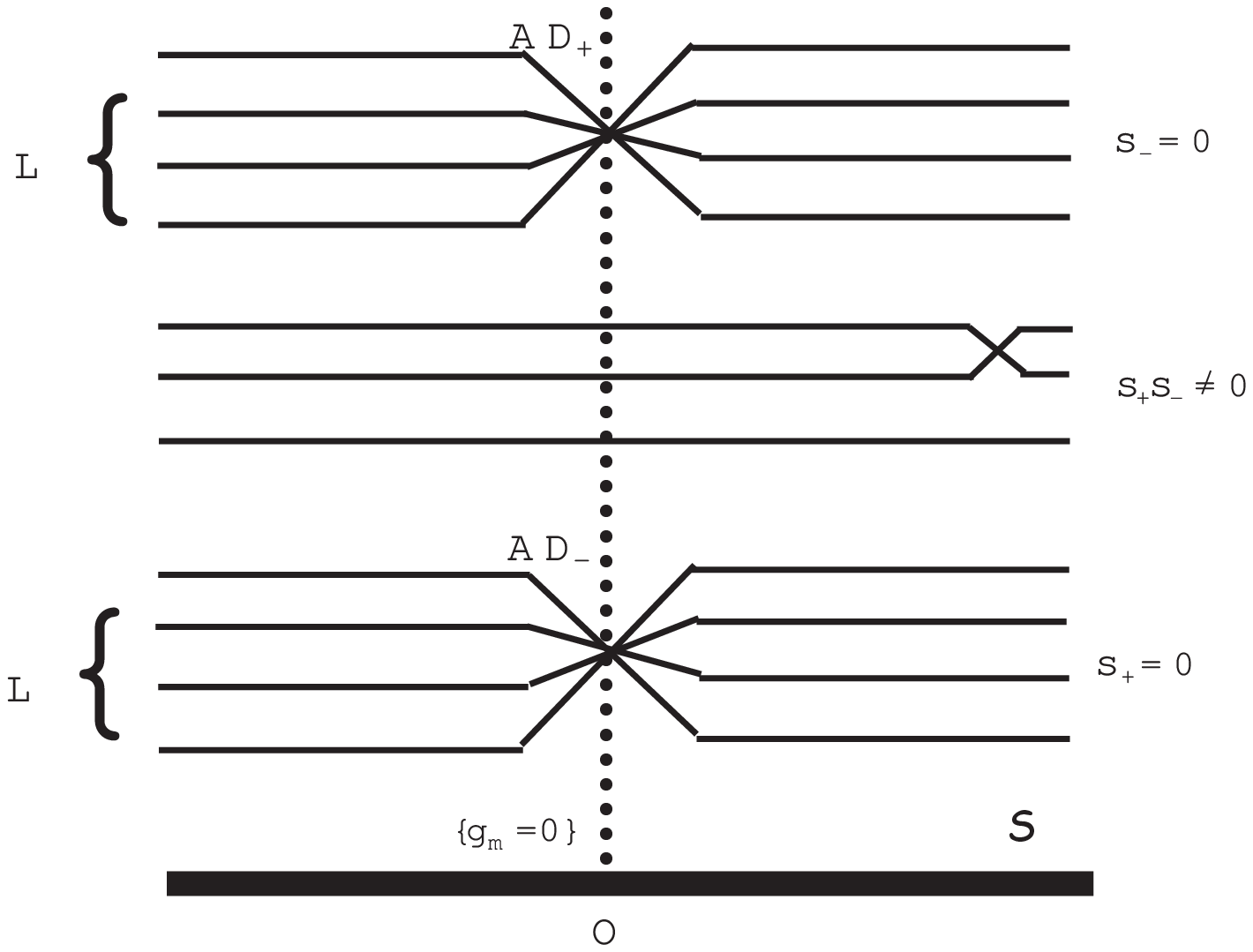}

\caption{{\em Features of the $\cN=1$ vacuum bundle
around the $\cN=1$ $AD$ points:} ~
The thick line at the bottom depicts  the coupling constant space $\cS$.
The $AD_+$, $AD_-$ branches are characterized by $s_-=0$, $s_+=0$
respectively. $AD_{\pm}$ points are on the fiber over (dotted line)
the origin $O$ of $\cS$ and are located at the $L$-fold ramification
points.
The remaining lines depict the branches with $s_+s_-\neq 0$.
}

\end{minipage}
\end{figure}


\newpage

\begin{figure}
\centering

\begin{minipage}{\linewidth}

\includegraphics[width=\linewidth]{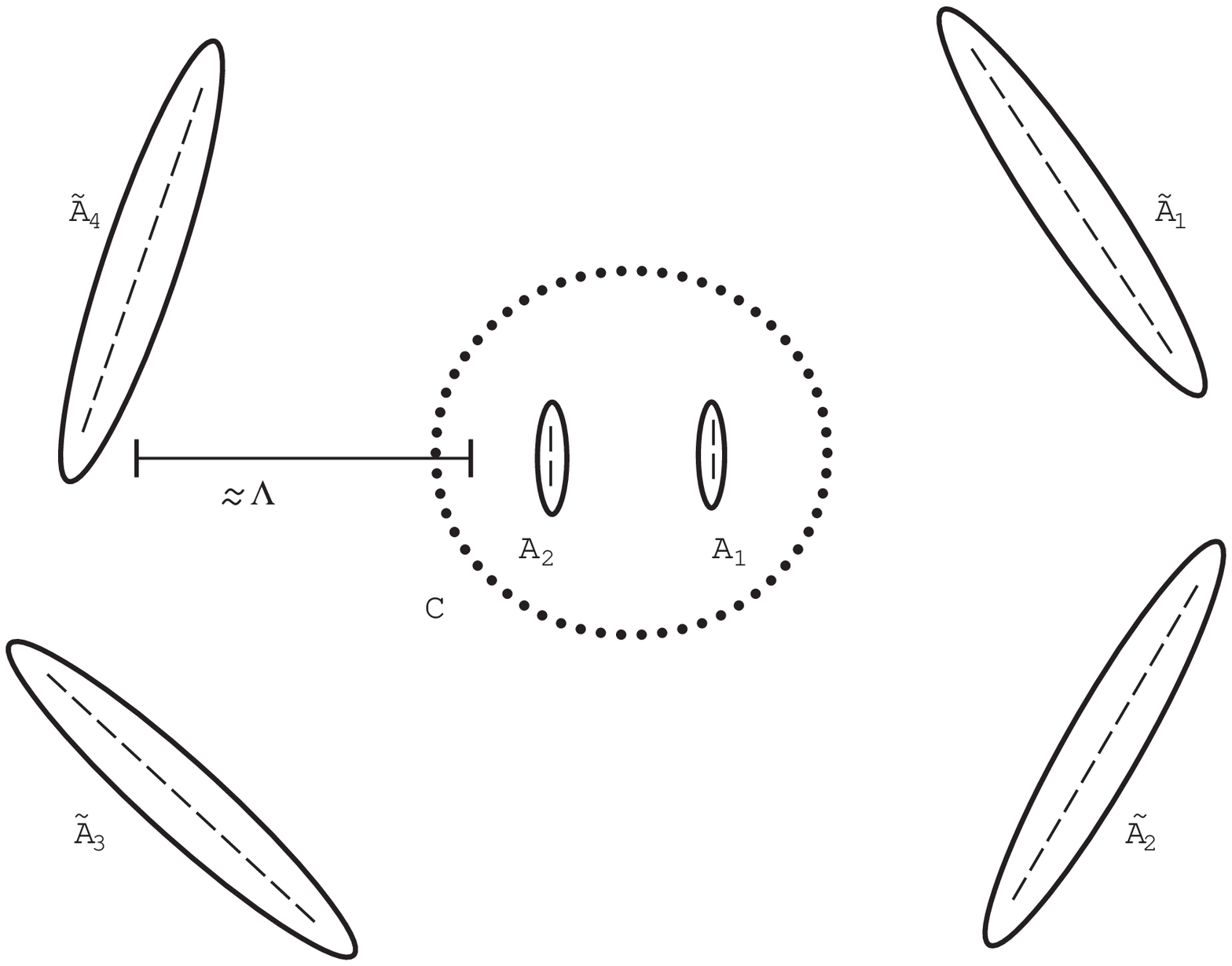}

\caption{{\em Cycles on the $\cN=1$ curve near the AD point 
in the case $N=8$, $n=6$:} ~
The broken lines depict the cuts. The ``long cycles'' $\tilde{A}_i$ 
correspond to the zeros of $H_{N-n}(x)^2F_{2n-N}(x)-4\La^{2N}$, 
while the  ``short cycles'' $A_i$ correspond to the zeros of $F_{2n-N}(x)$.
Latter cycles are relevant for the IR physics. 
The dotted circle $C$ defines the chiral operators in the 
strong coupling $\cN=1$ $SCFT_4$ by its contour integral. 
}

\end{minipage}
\end{figure}



\newpage

\begin{thebibliography}{99}



\bibitem{DV1}
R.~Dijkgraaf and C.~Vafa,
Nucl.\ Phys.\ B {\bf 644}, 3 (2002)
[arXiv:hep-th/0206255].

\bibitem{DV2}
R.~Dijkgraaf and C.~Vafa,
Nucl.\ Phys.\ B {\bf 644}, 21 (2002)
[arXiv:hep-th/0207106].

\bibitem{DV3}
R.~Dijkgraaf and C.~Vafa,
arXiv:hep-th/0208048.





\bibitem{Vafa}
C.~Vafa,
J.\ Math.\ Phys.\  {\bf 42}, 2798 (2001)
[arXiv:hep-th/0008142].

\bibitem{CIV}
F.~Cachazo, K.~A.~Intriligator and C.~Vafa,
Nucl.\ Phys.\ B {\bf 603}, 3 (2001)
[arXiv:hep-th/0103067].


\bibitem{CV}
F.~Cachazo and C.~Vafa,
arXiv:hep-th/0206017.



\bibitem{GV}
R.~Gopakumar and C.~Vafa,
Adv.\ Theor.\ Math.\ Phys.\  {\bf 3}, 1415 (1999)
[arXiv:hep-th/9811131].

\bibitem{OV1}
H.~Ooguri and C.~Vafa,
Nucl.\ Phys.\ B {\bf 577}, 419 (2000)
[arXiv:hep-th/9912123].


\bibitem{OV2}
H.~Ooguri and C.~Vafa,
Nucl.\ Phys.\ B {\bf 641}, 3 (2002)
[arXiv:hep-th/0205297].



\bibitem{DGLVZ}
R.~Dijkgraaf, M.~T.~Grisaru, C.~S.~Lam, C.~Vafa and D.~Zanon,
arXiv:hep-th/0211017.

\bibitem{CDSW}
F.~Cachazo, M.~R.~Douglas, N.~Seiberg and E.~Witten,
JHEP {\bf 0212}, 071 (2002)
[arXiv:hep-th/0211170].

\bibitem{Seiberg}
N.~Seiberg,
JHEP {\bf 0301}, 061 (2003)
[arXiv:hep-th/0212225].

\bibitem{CSW}
F.~Cachazo, N.~Seiberg and E.~Witten,
JHEP {\bf 0302}, 042 (2003)
[arXiv:hep-th/0301006].



\bibitem{CSW2}
F.~Cachazo, N.~Seiberg and E.~Witten,
arXiv:hep-th/0303207.

\bibitem{Konishi}
K.~Konishi,
Phys.\ Lett.\ B {\bf 135}, 439 (1984).
K.~Konishi and K.~Shizuya,
Nuovo Cim.\ A {\bf 90}, 111 (1985).




\bibitem{Feng}
B.~Feng,
arXiv:hep-th/0212274.

\bibitem{AO}
C.~Ahn and Y.~Ookouchi,
JHEP {\bf 0303}, 010 (2003)
[arXiv:hep-th/0302150].

\bibitem{Witten}
E.~Witten,
arXiv:hep-th/0302194.

\bibitem{BINOR}
A.~Brandhuber, H.~Ita, H.~Nieder, Y.~Oz and C.~Romelsberger,
arXiv:hep-th/0303001.

\bibitem{BFHN}
V.~Balasubramanian, B.~Feng, M.~x.~Huang and A.~Naqvi,
arXiv:hep-th/0303065.

\bibitem{NSW2}
S.~G.~Naculich, H.~J.~Schnitzer and N.~Wyllard,
arXiv:hep-th/0303268.

\bibitem{AC}
L.~F.~Alday and M.~Cirafici,
arXiv:hep-th/0304119.

\bibitem{CT}
R.~Casero and E.~Trincherini,
arXiv:hep-th/0304123.

\bibitem{KRS}
P.~Kraus, A.~V.~Ryzhov and M.~Shigemori,
arXiv:hep-th/0304138.




\bibitem{SW}
N.~Seiberg and E.~Witten,
Nucl.\ Phys.\ B {\bf 426}, 19 (1994)
[Erratum-ibid.\ B {\bf 430}, 485 (1994)]
[arXiv:hep-th/9407087];
N.~Seiberg and E.~Witten,
Nucl.\ Phys.\ B {\bf 431}, 484 (1994)
[arXiv:hep-th/9408099].



\bibitem{Ferrari}
F.~Ferrari,
Phys.\ Rev.\ D {\bf 67}, 085013 (2003)
[arXiv:hep-th/0211069];
Phys.\ Lett.\ B {\bf 557}, 290 (2003)
[arXiv:hep-th/0301157].




\bibitem{AD}
P.~C.~Argyres and M.~R.~Douglas,
Nucl.\ Phys.\ B {\bf 448}, 93 (1995)
[arXiv:hep-th/9505062].

\bibitem{APSW}
P.~C.~Argyres, M.~Ronen Plesser, N.~Seiberg and E.~Witten,
Nucl.\ Phys.\ B {\bf 461}, 71 (1996)
[arXiv:hep-th/9511154].

\bibitem{EHIY}
T.~Eguchi, K.~Hori, K.~Ito and S.~K.~Yang,
Nucl.\ Phys.\ B {\bf 471}, 430 (1996)
[arXiv:hep-th/9603002].


\bibitem{GVW}
S.~Gukov, C.~Vafa and E.~Witten,
Nucl.\ Phys.\ B {\bf 584}, 69 (2000)
[Erratum-ibid.\ B {\bf 608}, 477 (2001)]
[arXiv:hep-th/9906070].

\bibitem{ShV}
A.~D.~Shapere and C.~Vafa,
arXiv:hep-th/9910182.


\bibitem{GKP}
A.~Giveon, D.~Kutasov and O.~Pelc,
JHEP {\bf 9910}, 035 (1999)
[arXiv:hep-th/9907178].


\bibitem{GKM}
S.~Kharchev, A.~Marshakov, A.~Mironov, A.~Morozov and A.~Zabrodin,
Phys.\ Lett.\ B {\bf 275}, 311 (1992)
[arXiv:hep-th/9111037];
S.~Kharchev, A.~Marshakov, A.~Mironov, A.~Morozov and A.~Zabrodin,
Nucl.\ Phys.\ B {\bf 380}, 181 (1992)
[arXiv:hep-th/9201013];
S.~Kharchev, A.~Marshakov, A.~Mironov and A.~Morozov,
Nucl.\ Phys.\ B {\bf 397}, 339 (1993)
[arXiv:hep-th/9203043];
S.~Kharchev, A.~Marshakov, A.~Mironov and A.~Morozov,
Mod.\ Phys.\ Lett.\ A {\bf 8}, 1047 (1993)
[Theor.\ Math.\ Phys.\  {\bf 95}, 571 (1993\ TMFZA,95,280-292.1993)]
[arXiv:hep-th/9208046].


\bibitem{Pelc}
O.~Pelc,
JHEP {\bf 0003}, 012 (2000)
[arXiv:hep-th/0001054].


\bibitem{Seiberg-L}
N.~Seiberg,
Prog.\ Theor.\ Phys.\ Suppl.\  {\bf 102}, 319 (1990).
D.~Kutasov and N.~Seiberg,
Nucl.\ Phys.\ B {\bf 358}, 600 (1991).


\bibitem{ES2}
T.~Eguchi and Y.~Sugawara,
Nucl.\ Phys.\ B {\bf 598}, 467 (2001)
[arXiv:hep-th/0011148].


\bibitem{Losev}
A.~Losev,
Theor.\ Math.\ Phys.\  {\bf 95}, 595 (1993)
[Teor.\ Mat.\ Fiz.\  {\bf 95}, 307 (1993)]
[arXiv:hep-th/9211090].


\bibitem{EKYY}
T.~Eguchi, H.~Kanno, Y.~Yamada and S.~K.~Yang,
Phys.\ Lett.\ B {\bf 305}, 235 (1993)
[arXiv:hep-th/9302048];
T.~Eguchi, Y.~Yamada and S.~K.~Yang,
Mod.\ Phys.\ Lett.\ A {\bf 8}, 1627 (1993)
[arXiv:hep-th/9304121].


\bibitem{deBO}
J.~de Boer and Y.~Oz,
Nucl.\ Phys.\ B {\bf 511}, 155 (1998)
[arXiv:hep-th/9708044].


\bibitem{SW2}
A.~Hanany and Y.~Oz,
Nucl.\ Phys.\ B {\bf 452}, 283 (1995)
[arXiv:hep-th/9505075];
P.~C.~Argyres, M.~R.~Plesser and A.~D.~Shapere,
Phys.\ Rev.\ Lett.\  {\bf 75}, 1699 (1995)
[arXiv:hep-th/9505100];
J.~A.~Minahan and D.~Nemeschansky,
Nucl.\ Phys.\ B {\bf 464}, 3 (1996)
[arXiv:hep-th/9507032];
I.~M.~Krichever and D.~H.~Phong,
J.\ Diff.\ Geom.\  {\bf 45}, 349 (1997)
[arXiv:hep-th/9604199].


\bibitem{APS2}
P.~C.~Argyres, M.~Ronen Plesser and N.~Seiberg,
Nucl.\ Phys.\ B {\bf 471}, 159 (1996)
[arXiv:hep-th/9603042].



\bibitem{NSW}
S.~G.~Naculich, H.~J.~Schnitzer and N.~Wyllard,
JHEP {\bf 0301}, 015 (2003)
[arXiv:hep-th/0211254].




\bibitem{KSS}
D.~Kutasov, A.~Schwimmer and N.~Seiberg,
Nucl.\ Phys.\ B {\bf 459}, 455 (1996)
[arXiv:hep-th/9510222].


\bibitem{CM}
L.~Chekhov and A.~Mironov,
Phys.\ Lett.\ B {\bf 552}, 293 (2003)
[arXiv:hep-th/0209085].

\bibitem{CMMV}
L.~Chekhov, A.~Marshakov, A.~Mironov and D.~Vasiliev,
arXiv:hep-th/0301071.

\bibitem{IM}
H.~Itoyama and A.~Morozov,
Phys.\ Lett.\ B {\bf 555}, 287 (2003)
[arXiv:hep-th/0211259];
arXiv:hep-th/0301136.



\bibitem{TY}
S.~Terashima and S.~K.~Yang,
Phys.\ Lett.\ B {\bf 391}, 107 (1997)
[arXiv:hep-th/9607151];
Nucl.\ Phys.\ B {\bf 519}, 453 (1998)
[arXiv:hep-th/9706076].

\bibitem{GVY}
A.~Gorsky, A.~I.~Vainshtein and A.~Yung,
Nucl.\ Phys.\ B {\bf 584}, 197 (2000)
[arXiv:hep-th/0004087].


\bibitem{AGK}
R.~Auzzi, R.~Grena and K.~Konishi,
Nucl.\ Phys.\ B {\bf 653}, 204 (2003)
[arXiv:hep-th/0211282].







\end{thebibliography}
\end{document}